\documentclass{article}
\usepackage{amsmath,amsthm,amssymb,authblk}

\newtheorem{theorem}{Theorem}[section]

\newtheorem{lemma}[theorem]{Lemma}
\newtheorem{definition}[theorem]{Definition}
\newtheorem{example}[theorem]{Example}

\newtheorem{remark}[theorem]{Remark}
\numberwithin{equation}{section}
\def\pf{{\bf Proof.}~ }
\def\1.5{$1\frac{1}{2}$}

\begin{document}
\title{Quasi-cyclic Codes of Index $1\frac{1}{2}$}
\author{Yun Fan,\quad Hualu Liu\\
\small School of Mathematics and Statistics\\
\small Central China Normal University, Wuhan 430079, China}
\date{}
\maketitle

\insert\footins{\footnotesize{\it Email address}:
yfan@mail.ccnu.edu.cn (Yun Fan). hwlulu@aliyun.com (Hualu Liu).}

\begin{abstract}

We introduce quasi-cyclic codes of index $1\frac{1}{2}$,
construct such codes in terms of polynomials and matrices;
and prove that the quasi-cyclic codes of index $1\frac{1}{2}$ are asymptotically good.

\medskip
{\bf MSC classes:}~  94B05, 94B65, 15B52.

\end{abstract}

\section{Introduction}

Let $F$ be a finite field.
Any subspace $C$ of $F^n$ is called a linear code of length $n$ over $F$.
The fraction $R(C)=\frac{k}{n}$ is called the {\em rate} of $C$,
where $k=\dim C$ is the dimension of $C$.
The fraction $\Delta(C)=\frac{d}{n}$ is called the
{\em relative minimum distance} of $C$,
where $d=d(C)=\min_{0\ne{\bf c}\in C}{\rm w}({\bf c})$
is the minimum Hamming distance of $C$.
A sequence $C_1,C_2,\cdots$ of codes over $F$ with length
$n_i$ of $C_i$ going to infinity is said to be {\em asymptotically good}
if both the rate $R(C_i)$ and the relative minimum distance
$\Delta(C_i)$ are positively bounded from below.
A class of codes is said to be {\em asymptotically good} if there exist
asymptotically good sequences of codes within the class.
By a Varshamov's random argument \cite{V}, linear codes
are asymptotically good.

Let $A$ be a permutation group on the the index set $\{1,2,\cdots,m\}$
of coordinates of $F^m$.
If a linear code $C$ in $F^m$ is invariant by the $A$-action,
i.e. $\alpha({\bf c})\in C$, $\forall$ ${\bf c}\in C$ $\forall$ $\alpha\in A$,
then $C$ is said to be an {\em $A$-acted code} (\cite{BM}), or
an {\em $A$-permutation code} (\cite{FY}).
If $A=\langle (12\cdots m)\rangle$ is a cyclic group generated by
the cycle $(12\cdots m)$ and $C$ is an $A$-acted code,
then $C$ is just the so-called cyclic code of length $m$.
It is a long standing open question (see \cite{MW06}):
{\em whether or not the cyclic codes are asymptotically good?}

Consider the vector space $F^m\times F^m$ over $F$
with $\{1,2,\cdots,m\}$ and $\{m+1,m+2,\cdots,2m\}$, respectively,
being the index sets of coordinates of the first $F^m$ and the second $F^m$, respectively.
Let $A=\langle (12\cdots m)(m+1,m+2,\cdots,2m)\rangle$ be the permutation
group generated by the product of the corresponding
two cycle $(12\cdots m)$ and $(m+1,m+2,\cdots,2m)$ of length $m$.
Then $A$ is a cyclic group of order $m$; and
any $A$-acted code $C$ in $F^m\times F^m$ is said to be
a {\em quasi-cyclic code} of {\em index $2$} and {\em co-index $m$}.
Similarly, {\em quasi-cyclic codes} of {\em index $n$ and co-index $m$}
are defined to be the subspaces $C$ of $F^m\times\cdots\times F^m$ (with $n$ copies)
which are invariant by the permutation which is the product
of $n$ disjoint cycles of length $m$.
Specifically, the quasi-cyclic codes of index $1$ and co-index $m$
are just the cyclic codes of length $m$.

By a random method, \cite{CPW} showed that,
if $2$ is primitive for infinitely many primes
(this is a so-called {\em Artin's conjecture}),
then asymptotically good binary quasi-cyclic codes of index $2$ exist.
Later, \cite{C} and \cite{K} made big improvements
%from different points of view
and proved that, without the Artin's conjecture,
the binary quasi-cyclic codes of index $2$ are asymptotically good.
On the other hand, Ling and Sol\'e \cite{LS} showed that self-dual quasi-cyclic
codes of co-index $3$ and index going to infinity are asymptotically good.

For arbitrary finite group $G$ of order $m$,
any left ideal $C$ of the group ring $FG$ is called a {\em group code};
and any $FG$-submodule of $(FG)^n$ is called
a {\em quasi-group code} of {\em index $n$} and {\em co-index $m$}.
If $G$ is abelian, then the quasi-group codes are also called
{\em quasi-abelian codes}, see \cite{DR, W}.
Quasi-abelian codes are just quasi-cyclic codes once $G$ is cyclic.

%In 2006,
Bazzi and Mitter \cite{BM} obtained by a random method
two asymptotically good classes of codes:
(1) binary quasi-abelian codes of index $2$;
(2) binary dihedral group codes.
Soon after, with the similar random method,
Mart\'inez-P\'erez and Willems \cite{MW} showed that
self-dual doubly-even binary dihedral group codes are asymptotically good.
By a result \cite[Theorem 3.3]{FL},
the two asymptotically good classes of binary codes obtained in \cite{BM}
can be extended to any $q$-ary case.
It was also shown in \cite{FL} that the quasi-abelian codes
with index going to infinity are asymptotically good.
The dihedral groups are the non-abelian finite groups which are nearest to
cyclic groups. However,
if the actions of the involutions (elements of order $2$) of the
dihedral groups are ignored, then the dihedral group codes can be viewed
as quasi-cyclic codes of index $2$.

Thus an interesting question we are concerned with comes up:
is it possible to consider the quasi-cyclic codes of fractional index between $1$ and $2$?
If it is, are such codes asymptotically good?
%whether there are any asymptotically good classes of quasi-cyclic codes
%between quasi-cyclic codes of index $1$ (i.e., cyclic codes)
%and quasi-cyclic codes of index $2$?

In this paper we introduce {\em quasi-cyclic codes of index $1\frac{1}{2}$},
and show that such codes are asymptotically good.

In Section 2, we define the quasi-cyclic codes of index $1\frac{1}{2}$
and co-index $2m$ to be the subspaces of $F^{2m}\times F^m$
which are invariant by a permutation which is a product  of
two disjoint cycles of length~$2m$ and length~$m$, respectively
(hence the permutation generates a cyclic group of order $2m$);
and construct such codes in terms of polynomials and matrices.

In section 3, we study a kind of random quasi-cyclic codes of index $1\frac{1}{2}$.
We exhibit, in an asymptotic and probabilistic sense, a positive lower bound
of the relative minimal distances of such random codes,
see Theorem \ref{Delta C} below.
Then it follows that asymptotically good quasi-cyclic codes of index $1\frac{1}{2}$
exist, see Theorem \ref{a g} below.

\section{Quasi-cyclic codes of index $1\frac{1}{2}$}

From now on, we always assume that $F$ is a finite field with $q$ elements,
where $q$ is a power of an odd prime;
and $m$ is a positive integer coprime to $q$.
For fundamentals on finite rings and coding theory,
please refer to \cite{HP, Mc}.

By $R_{2m}=F[X]/\langle X^{2m}-1\rangle$ we denote the residue ring
of the polynomial ring $F[X]$ over $F$ modulo the ideal
$\langle X^{2m}-1\rangle$ generated by $X^{2m}-1$.
Similarly, $R_{m}=F[X]/\langle X^{m}-1\rangle$.
Consider the product
$$ R_{2m}\times R_m
    =F[X]/\langle X^{2m}-1\rangle \times F[X]/\langle X^{m}-1\rangle.
$$
Each element of $R_{2m}\times R_m $ is uniquely represented as
$$\big(a(X),a'(X)\big)\quad {\rm with}\quad
a(X)=\sum_{j=0}^{2m-1}a_{j}X^{j},~a'(X)=\sum_{j'=0}^{m-1}a'_{j'}X^{j'}\in F[X].
$$
We always identify the element $\big(a(X),a'(X)\big)\in R_{2m}\times R_m$
with the word
$$(a_{0},a_{1},\cdots,a_{2m-2},a_{2m-1},~a'_{0},a'_{1},\cdots,a'_{m-2},a'_{m-1})
 \in F^{2m}\times F^m. $$

Let $\xi$ be a permutation of the coefficients of $F^{2m}\times F^m$,
which is a product of two disjoint cycles of length $2m$ and $m$, respectively,
as follows:
\begin{eqnarray*}
&&
\xi(a_{0},a_{1},\cdots,a_{2m-2},a_{2m-1},~a'_{0},a'_{1},\cdots,a'_{m-2},a'_{m-1})\\
&&=(a_{2m-1},a_{0},a_1,\cdots,a_{2m-2},~
 a'_{m-1},a'_{0},a'_1,\cdots,a'_{m-2}).
\end{eqnarray*}
The permutation $\xi$ on $F^{2m}\times F^m$
is corresponding to the operation on $R_{2m}\times R_m$ by multiplying $X$:
%as follows:
for $a(X)=\sum_{j=0}^{2m-1}a_{j}X^{j}$
and $a'(X)=\sum_{j'=0}^{m-1}a'_{j'}X^{j'}$,
\begin{equation}\label{by X}
X\big(a(X),a'(X)\big)=
\big(Xa(X)~({\rm mod}~X^{2m}-1),~ Xa'(X)~({\rm mod}~X^{m}-1)\big).
\end{equation}

\begin{definition}\label{qc 1.5}\rm
A linear subspace $C$ of $R_{2m}\times R_m$ is said to be a {\em quasi-cyclic code over $F$
of index $1\frac{1}{2}$  and co-index $2m$} if $C$ is invariant
by the permutation~$\xi$, i.e.
\begin{equation*}
 X\big(c(X),\,c'(X)\big)\in C,\qquad \forall~~\big(c(X),c'(X)\big)\in C.
\end{equation*}
%at that case we denote $C\le R_{2m}\times R_m$.
\end{definition}

\smallskip
The operation \eqref{by X} can be extended in a natural way: for any $f(X)\in F[X]$
and any $\big(a(X),a'(X)\big)\in R_{2m}\times R_m$,
$$\begin{array}{l}
f(X)\big(a(X),a'(X)\big)\\[5pt]
=\big(f(X)a(X)~({\rm mod}~X^{2m}-1),\, f(X)a'(X)~({\rm mod}~X^{m}-1)\big).
\end{array}
\eqno(\ref{by X}')
$$
To shorten the notation, in the following we abbreviate the operation (\ref{by X}$'$)
on $R_{2m}\times R_m$ as:
$$f(X)\big(a(X),a'(X)\big)=\big(f(X)a(X),f(X)a'(X)\big).$$
Obviously, a linear subspace $C$ of $R_{2m}\times R_m$ is a
quasi-cyclic code of index $1\frac{1}{2}$ and co-index $2m$
if and only if it is invariant by multiplying any $f(X)\in F[X]$.
In other words, $R_{2m}\times R_m$ is an $F[X]$-module (or, $R_{2m}$-module),
and its $F[X]$-submodules ($R_{2m}$-modules)
are just the quasi-cyclic codes of index $1\frac{1}{2}$
and co-index $2m$.

An $F[X]$-submodule of $R_{2m}\times R_m$ is generated by at most two elements.
For our later use, we illustrate a kind of quasi-cyclic codes
of index $1\frac{1}{2}$ and co-index $2m$,
each of which is generated by one element.

\begin{example}\label{class 1}\rm
For $\big(a(X),a'(X)\big)\in R_{2m}\times R_m$, let
\begin{equation}\label{C_a,a'}
C_{a,a'}=\big\{\big(f(X)a(X),f(X)a'(X)\big)\in R_{2m}\times R_m
 \,\big|\,f(X)\in R_{2m}\big\}.
\end{equation}
Then $C_{a,a'}$ is a quasi-cyclic code of index $1\frac{1}{2}$ and co-index $2m$.
Further, let
$$\begin{array}{l}
a(X)=a_0+a_1X+\cdots+ a_{2m-1}X^{2m-1},\\[5pt]
a'(X)=a'_0+a'_1X+\cdots+a'_{m-1}X^{m-1}.
\end{array}$$
For $a(X)$ we have a $2m$-dimensional vector $(a_0,a_1,\cdots,a_{2m-2},a_{2m-1})$,
from which a circulant $2m\times 2m$ matrix is constructed as follows:
$$
A=\begin{pmatrix}a_0 & a_1 & \cdots & a_{2m-2} & a_{2m-1} \\
 a_{2m-1} & a_0 &\cdots &a_{2m-3}& a_{2m-2} \\
 \cdots & \cdots & \cdots & \cdots & \cdots\\
 a_{2} & a_{3} & \cdots & a_0& a_{1}\\
 a_{1}  & a_{2} & \cdots & a_{2m-1} & a_0
\end{pmatrix}_{2m\times 2m}.
$$
Similarly, from $a'(X)$ we have a circulant $m\times m$ matrix as follows:
$$
A'=\begin{pmatrix}a'_0 & a'_1 & \cdots & a'_{m-2} & a'_{m-1} \\
 a'_{m-1} & a'_0 &\cdots &a'_{m-3}& a'_{m-2} \\
 \cdots & \cdots & \cdots & \cdots & \cdots\\
 a'_{2} & a'_{3} & \cdots & a'_0& a'_{1}\\
 a'_{1}  & a'_{2} & \cdots & a'_{m-1} & a'_0
\end{pmatrix}_{m\times m}.
$$
Then we can construct a $2m\times 3m$ matrix:
\begin{equation}
 \widehat A=\left(\begin{array}{cc}
  \mbox{\Huge A} & \begin{array}[b]{c} A'\\ A' \end{array}
 \end{array}\right)_{2m\times 3m}.
\end{equation}
And it is easy to see that
$$
C_{a,a'}=
\big\{(y_{0},y_1\cdots,y_{2m-1})\widehat A\in F^{2m}\times F^m
    \,\big|\,(y_{0},y_1\cdots,y_{2m-1})\in F^{2m}\big\}.
$$
Note that $\widehat A$ is not a generator matrix of the code $C_{a,a'}$ in general,
since it is not of rank~$2m$ in general. After Theorem \ref{thm2} below,
we'll see how to get a generator matrix of $C_{a,a'}$ from the matrix $\widehat A$.
\end{example}

\begin{remark}\label{rem CRT}
\rm For $X^{2m}-1$, we have the following facts:
\begin{equation*}
X^{2m}-1=(X^m+1)(X^m-1),~~~~
\frac{1}{2}(X^m+1) + \frac{-1}{2}(X^m-1)=1.
\end{equation*}
By Chinese Remainder Theorem, we have the natural isomorphism:
\begin{equation}\label{CRT}
\begin{array}{cccc}
\frac{F[X]}{\langle X^{2m}-1\rangle}&\mathop{\longrightarrow}\limits^{\cong}&
\frac{F[X]}{\langle X^{m}-1\rangle}\times\frac{F[X]}{\langle X^{m}+1\rangle},
\\[8pt]
 f(X) & \longmapsto &
  \big(f(X)~({\rm mod}~X^m-1),~f(X)~({\rm mod}~X^m+1)\big).
\end{array}
\end{equation}
In the following, for any polynomial $f(X)$, by
$\big\langle f(X)\big\rangle_{R_{2m}}$ we clarify that it is an ideal
of $R_{2m}$ generated by $f(X)$.
It is easy to check that the isomorphism \eqref{CRT} induces a direct sum:
$$R_{2m}=\big\langle X^m+1\big\rangle_{R_{2m}}
 \oplus \big\langle X^m-1\big\rangle_{R_{2m}},$$
and two natural isomorphisms:
\begin{equation}\label{R+-}
\begin{array}{ccc}
\big\langle X^m+1\big\rangle_{R_{2m}}
&\mathop{\longrightarrow}\limits^{\cong}&F[X]/\langle X^m-1\rangle,
\\
\big\langle X^m-1\big\rangle_{R_{2m}}
&\mathop{\longrightarrow}\limits^{\cong}&F[X]/\langle X^m+1\rangle.
\end{array}
\end{equation}
\end{remark}

\smallskip
\begin{theorem}\label{thm2}
Given any $(a(X),a'(X))\in R_{2m}\times R_m$.
Let
\begin{equation}\label{g_a,a'}
 g_{a,a'}(X)=\gcd\big(a(X),X^m+1\big)\cdot\gcd\big(a(X),a'(X),X^m-1\big)
\end{equation}
where $\gcd(\cdots)$ denotes the greatest common divisor, and let
$$h_{a,a'}(X)=\frac{X^{2m}-1}{g_{a,a'}(X)}.$$
Then $(a(X),a'(X))$ induces an $F[X]$-homomorphism
$$\gamma_{a,a'}:
 R_{2m}\longrightarrow R_{2m}\times R_m,\quad 
 f(X)\longmapsto \big(f(X)a(X),\;f(X)a'(X)\big),
$$
and the following hold:
\begin{itemize}
\item[(i)]
 The image ${\rm im}(\gamma_{a,a'})=C_{a,a'}$, where $C_{a,a'}$ is defined
 in Eqn \eqref{C_a,a'}.
\item[(ii)]
 The kernel ${\rm ker}(\gamma_{a,a'})
=\langle h_{a,a'}(X)\rangle_{R_{2m}}$,
hence $\dim C_{a,a'}=\deg h_{a,a'}(X)$.
\item[(iii)] $\gamma_{a,a'}$ induces an isomorphism
$\langle g_{a,a'}(X)\rangle_{R_{2m}}
\mathop{\longrightarrow}\limits^{\cong} C_{a,a'}$; in particular,
$$\kern-3mm 
  C_{a,a'}=\big\{\big(b(X)a(X),b(X)a'(X)\big)\in R_{2m}\times R_m\,\big|\,
   b(X)\in\langle g_{a,a'}(X)\rangle_{R_{2m}}\big\}.
$$
\end{itemize}
\end{theorem}

\pf It is obvious that $\gamma_{a,a'}$ is an $F[X]$-homomorphism and (i) holds.

For (ii), $f(X)\in{\rm ker}(\gamma_{a,a'})$ if and only if
\begin{equation}\label{system}
\left\{\begin{matrix}f(X)a(X)\equiv 0\pmod{X^{2m}-1},\\
 f(X)a'(X)\equiv 0\pmod{X^{m}-1}.\end{matrix}\right.
\end{equation}
By the isomorphism \eqref{CRT},
the system \eqref{system} is equivalent to the following system:
$$\left\{\begin{array}{ll}f(X)a(X)\equiv 0&\pmod{X^{m}+1},\\
f(X)a(X)\equiv 0&\pmod{X^{m}-1},\\
 f(X)a'(X)\equiv 0&\pmod{X^{m}-1}.\end{array}\right.$$
The last two equations are combined into one equation:
$$f(X)\gcd(a(X),a'(X))\equiv 0~({\rm mod}~X^m-1).$$
So the system \eqref{system} is equivalent to:
$$\left\{\begin{array}{ll}f(X)\equiv 0
   & \big({\rm mod}~{\frac{X^{m}+1}{\gcd(a(X),X^m+1)}}\big),\\[8pt]
 f(X)\equiv 0 &\big({\rm mod}~\frac{X^{m}-1}{\gcd(a(X),a'(X),X^m-1)}\big).
 \end{array}\right.$$
Since $\frac{X^{m}+1}{\gcd(a(X),X^m+1)}$ and
$\frac{X^{m}-1}{\gcd(a(X),a'(X),X^m-1)}$ are coprime to each other, we see that
the system \eqref{system} holds if and only if
$$\textstyle
f(X)\equiv 0\quad \big({\rm mod}~\frac{X^{m}+1}{\gcd(a(X),X^m+1)}\cdot
\frac{X^{m}-1}{\gcd(a(X),a'(X),X^m-1)}\big),
$$
that is, $f(X)\in \langle h_{a,a'}(X)\rangle_{R_{2m}}$. In particular,
$$\dim C_{a,a'}=\dim R_{2m}-\dim{\rm ker}(\gamma_{a,a'})
=2m-\deg g_{a,a'}(X)=\deg h_{a,a'}(X).$$
We are done for (ii).

Finally, since $X^{2m}-1$ has no multiple roots (i.e., $R_{2m}$ is semisimple),
$$R_{2m}=\langle h_{a,a'}(X)\rangle_{R_{2m}}\oplus
\langle g_{a,a'}(X)\rangle_{R_{2m}}.$$
And the kernel of the homomorphism $\gamma_{a,a'}$ is just
the ideal $\langle h_{a,a'}(X)\rangle_{R_{2m}}$.
So, (iii) is proved.
\qed

\begin{example}\rm
 Take $q=3$, $m=2$, $a(X)=(X-1)(X^2+1)=X^3+2X^2+X+2$
and $a'(X)=X+1$. By Theorem \ref{thm2},
$$\begin{array}{ll}
g_{a,a'}(X)=\gcd\big(a(X),X^2+1\big)\gcd\big(a(X),a'(X),X^2-1\big)=X^2+1;\\[5pt]
 h_{a,a'}(X)=(X^{4}-1)/g_{a,a'}(X)=X^2-1.
\end{array}$$
So $\dim C_{a,a'}=2$. Using the notations in Example \ref{class 1}, we have
$$
 A=\begin{pmatrix}2&1&2&1\\ 1&2&1&2\\ 2&1&2&1\\ 1&2&1&2 \end{pmatrix},
 \qquad A'=\begin{pmatrix}1&1\\1 &1 \end{pmatrix};
$$
$$
\widehat A=\begin{pmatrix}2&1&2&1&1&1\\ 1&2&1&2&1&1\\
 2&1&2&1&1&1\\ 1&2&1&2&1&1 \end{pmatrix}.
$$
Thus the first two rows of $\widehat A$ are linearly independent and
$$
 G=\begin{pmatrix} 2&1&2&1&1&1\\1&2&1&2&1&1\end{pmatrix}
$$
is a generator matrix of $C_{a,a'}$. The $9$ codewords of $C_{a,a'}$ are
as follows:
$$\begin{array}{ccccc} 
000000 & 212111 & 121222 & 121211 & 212122 \\
000022 & 000011 & 121200 & 212100
\end{array}$$
\end{example}

\begin{example}\rm
 Take $q=3$, $m=2$, $a(X)=X-1=X+2$
and $a'(X)=X-1$. By Theorem \ref{thm2},
$$\begin{array}{ll}
g_{a,a'}(X)=\gcd\big(a(X),X^2+1\big)\gcd\big(a(X),a'(X),X^2-1\big)=X-1;\\[5pt]
h_{a,a'}(X)=(X^{4}-1)/g_{a,a'}(X)=(X^2+1)(X+1).
\end{array}$$
So $\dim C_{a,a'}=3$. And by the notations in Example \ref{class 1},
$$
 A=\begin{pmatrix}2&1\\ &2&1\\ &&2&1\\ 1&&&2 \end{pmatrix},
 \qquad A'=\begin{pmatrix}2&1\\ 1&2 \end{pmatrix};
$$
$$
\widehat A=\begin{pmatrix}2&1&&&2&1\\ &2&1&&1&2\\
 &&2&1&2&1\\ 1&&&2&1&2 \end{pmatrix}.
$$
Thus the first three rows of $\widehat A$ are linearly independent and
$$
 G=\begin{pmatrix}2&1&&&2&1\\ &2&1&&1&2\\&&2&1&2&1\end{pmatrix}
$$
is a generator matrix of $C_{a,a'}$. The $27$ codewords of $C_{a,a'}$ are
as follows.
$$\begin{array}{ccccccc} 
000000 & 002121 & 001212 & 021012 & 020100 & 022221 & 012021 \\
011112 & 010200 & 210021 & 212112 & 211200 & 201000 & 200121 \\
202212 & 222012 & 221100 & 220221 & 120012 & 122100 & 121221 \\
111021 & 110112 & 112200 & 102000 & 101121 & 100212 &  \\
\end{array}$$
\end{example}

From Example~\ref{class 1}, we construct a class of quasi-cyclic codes of
index $1\frac{1}{2}$ and co-index $2m$, which will be used in the next section.

\begin{example}\label{class 11}\rm
Consider an ideal $J^+_{2m}$ of $R_{2m}$ and an ideal $J_m$ of $R_m$ as follows:
\begin{equation}\label{J}
{J}^+_{2m}=\big\langle (X^m+1)(X-1)\big\rangle_{R_{2m}},\qquad
{J}_{m}=\langle X-1\rangle_{R_{m}}.
\end{equation}
For any $a(X)\in J^+_{2m}$ and $a'(X)\in J_m$,
as illustrated in Example \ref{class 1}, the code
$C_{a,a'}=\{(f(X)a(X),f(X)a'(X))\in R_{2m}\times R_m\,|\,f(X)\in R_{2m}\}$
is a quasi-cyclic code of index $1\frac{1}{2}$ and co-index $2m$.
By the definition of $J^+_{2m}$ and $J_m$ in Eqn~\eqref{J},
$$
 \gcd(a(X),X^m+1)=X^m+1,\qquad (X-1)\,|\,\gcd(a(X),a'(X),X^m-1).
$$
So $(X^m+1)(X-1)\,|\,g_{a,a'}(X)$, where
$
g_{a,a'}(X)%=\gcd(a(X),X^m+1)\cdot\gcd(a(X),a'(X),X^m-1)
$
is defined in Eqn \eqref{g_a,a'}. Then
$$
\big\langle g_{a,a'}(X)\big\rangle_{R_{2m}}\subseteq
\big\langle (X^m+1)(X-1)\big\rangle_{R_{2m}}=J^+_{2m}.
$$
By Theorem \ref{thm2} (iii), instead of $R_{2m}$,
the quasi-cyclic code $C_{a,a'}$ of index $1\frac{1}{2}$ and co-index $2m$
can be formed within $J^+_{2m}$ as follows:
\begin{equation}\label{JC_a,a'}
C_{a,a'}=\big\{(b(X)a(X),b(X)a'(X))\in R_{2m}\times R_m
 \,\big|\,b(X)\in J^+_{2m}\big\}.
\end{equation}
\end{example}

\begin{remark}\label{J_m}\rm
 Let $\frac{X^m-1}{X-1}=p_1(X)\cdots p_h(X)$
be an irreducible decomposition of $\frac{X^m-1}{X-1}$ in $F[X]$,
i.e., all the $p_j(X)$'s are irreducible polynomials over $F$.
For $j=1,\cdots,h$, by Eqn \eqref{R+-}, the following is a surjective homomorphism:
$$\mu_{2m}^{(j)}:~~ J_{2m}^+\to F[X]/\langle p_j(X)\rangle,~~
 a(X)\mapsto a(X)~({\rm mod}~p_j(X)).$$
And, by Chinese Remainder Theorem,
we have an isomorphisms as follows.
\begin{equation}\label{mu_2m}
\begin{array}{cccc}
\mu_{2m}:& J_{2m}^+&\to& F[X]/\langle p_1(X)\rangle\times\cdots
  \times F[X]/\langle p_h(X)\rangle , \\
 & a(X) &\mapsto& \Big(\,\mu_{2m}^{(1)}\big( a(X)\big),~\cdots,~
 \mu_{2m}^{(h)}\big( a(X)\big)\,\Big).
\end{array}
\end{equation}
Similarly, for $J_m$ we have surjective homomorphisms as follows.
$$\mu_{m}^{(j)}:~ J_{m}\to F[X]/\langle p_j(X)\rangle,~
 a'(X)\mapsto a'(X)~({\rm mod}~p_j(X)); \quad j=1,\cdots,h.$$
And an isomorphisms is as follows.
$$\begin{array}{cccc}
\mu_{m}:& J_{m}&\to& F[X]/\langle p_1(X)\rangle\times\cdots
  \times F[X]/\langle p_h(X)\rangle , \\
 & a'(X) &\mapsto& \Big(\,\mu_{m}^{(1)}\big( a'(X)\big),~\cdots,~
 \mu_{m}^{(h)}\big( a'(X)\big)\,\Big).
\end{array}$$
In particular, $\dim J_{2m}^+=\dim J_m=m-1$.
\end{remark}

\begin{lemma}\label{dC_a,a'}
Let notations be as in Remark \ref{J_m},
let $\big(a(X),a'(X)\big)\in J^+_{2m}\times J_m$. Then
$\dim C_{a,a'}\le m-1$; and $\dim C_{a,a'}<m-1$
if and only if there is an index~$j$ with $1\le j\le h$ such that
$\mu_{2m}^{(j)}(a(X))=0=\mu_{m}^{(j)}(a'(X))$.
\end{lemma}

\pf From the assumption of the lemma and the
isomorphisms $\mu_{2m}$ and $\mu_m$ in Remark \ref{J_m},
it is easy to see that $\gcd\big(a(X),X^m+1\big)=X^m+1$ and
$$
\gcd\big(a(X),a'(X), X^m-1\big)=(X-1)
  \prod_{\mu_{2m}^{(j)}(a(X))=0=\mu_{m}^{(j)}(a'(X))}p_j(X).
$$
The conclusions follow from Theorem \ref{thm2}. \qed

\begin{lemma}\label{number I}
Let notations be as in Remark \ref{J_m}, let
\begin{equation}\label{l_m}
\ell_m=\min\{\deg p_1(X),\cdots,\deg p_h(X)\}.
\end{equation}
Then any non-zero ideal of $R_{2m}$ which are contained in $J^+_{2m}$
has dimension at least $\ell_m$, and the number of the ideals 
of dimension $d$ (with $\ell_m\le d<m$) which are contained in $J^+_{2m}$
is at most $m^{\frac{d}{\ell_m}}$.
\end{lemma}

\pf By the isomorphism \eqref{mu_2m}, 
each irreducible ideal contained in $J^+_{2m}$ is corresponding 
to exact one irreducible divisor of $\frac{X^m-1}{X-1}$ such that 
the dimension of the ideal is equal to the degree of the corresponding divisor.
Thus, the minimal dimension of the ideals contained in $J^+_{2m}$ 
is equal to $\ell_m$. And, any $d$-dimensional ideal contained in $J^+_{2m}$
is a sum of at most $d/\ell_m$ irreducible ideals.
So the number of the $d$-dimensional ideals contained in $J^+_{2m}$
is at most the partial sum of binomial coefficients 
$\sum_{i=1}^{\lfloor d/\ell_m\rfloor}{h\choose i}$,
where $h$ is the number of the irreducible ideals contained in $J^+_{2m}$
(as in Remark \ref{J_m}) and $\lfloor d/\ell_m\rfloor$ denotes 
the largest integer which is not larger than $d/\ell_m$.
It is easy to check that the partial sum is not larger than $m^{d/\ell_m}$.
\qed

\section{Random quasi-cyclic codes of index $1\frac{1}{2}$}

Keep the notations in Section 2.

Let $h_q(x)=x\log_q(q-1)-x\log_q x-(1-x)\log_q(1-x)$
with the convention that $0\log_q0=0$,
it is called the {\em $q$-ary entropy}.
Note that $h_q(x)$ is a strictly increasing concave function
in the interval $[0,1-q^{-1}]$ with $h_q(0)=0$ and $h_q(1-q^{-1})=1$,
see \cite[\S 2.10.6]{HP}.
Hence, in the interval $[0,1]$, the inverse function $h_q^{-1}(x)$
exists and it is a strictly increasing convex function
with $h_q^{-1}(0)=0$ and $h_q^{-1}(1)=1-q^{-1}$.
Specifically,  $h_q^{-1}(1/2)< 1/2$.

In the following, we always assume that $\delta$ is a positive real number less than
$\frac{2}{3}h_q^{-1}(\frac{1}{2})$; i.e.,
\begin{equation}\label{delta}\textstyle
0<\delta<\frac{2}{3}h_q^{-1}(\frac{1}{2})<\frac{1}{3}.
\end{equation}

Let $J^+_{2m}=\big\langle(X^m+1)(X-1)\big\rangle_{R_{2m}}$
and $J_m=\langle X-1\rangle_{R_{m}}$ as
in Eqn \eqref{J} of Example \ref{class 11}.
In this section we view the set $J^+_{2m}\times J_m$
as a probability space, whose samples are afforded with equal probability.
We will study a kind of random quasi-cyclic codes of index $1\frac{1}{2}$
over the probability space. %% from Example~\ref{class 11}.
As a preparation, we introduce a type of $0$-$1$ variables.

Given any $b(X)\in J^+_{2m}$. We define a Bernoulli variable $X_b$
over the probability space $J^+_{2m}\times J_m$ as follows:
for all samples $\big(a(X),a'(X)\big)\in J^+_{2m}\times J_m$,
\begin{equation}\label{X_b}
X_b=\begin{cases} 1, & 1\le {\rm w}\big(b(X)a(X), b(X)a'(X)\big)\le 3m\delta;\\
 0, & {\rm otherwise}.\end{cases} \quad
\end{equation}
Since $b(X)\in J^+_{2m}$, the set
$\{b(X)a(X)\in R_{2m}\,|\,a(X)\in J^+_{2m}\}$ is the ideal of $R_{2m}$
generated by $b(X)$, we denote it by $I_b$ and denote its dimension by $d_b$; i.e.,
\begin{equation}\label{I_b}
 I_b=\big\langle b(X)\big\rangle_{R_{2m}}\subseteq J^+_{2m}, \quad d_b=\dim I_b.
\end{equation}
In $R_m$, since $b(X)\in J_m$, we get an ideal
$$
 I'_b=\big\langle b(X)\big\rangle_{R_{m}}
 =\{b(X)a'(X)\in R_{m}\,|\,a'(X)\in J_{m}\} \subseteq J_{m}.
$$
Note that $\dim I'_b=\dim I_b=d_b$ (cf. Remark \ref{J_m}).
We show an estimation of the expectation ${\rm E}(X_b)$
of the random variable $X_b$ for later quotation.

\begin{lemma}\label{EX_b}
 Let notations be as in Eqns \eqref{delta}-\eqref{I_b}. Then
$$ {\rm E}(X_b)\le q^{-2d_b+2d_bh_q(\frac{3}{2}\delta)+\log_q m}. $$
\end{lemma}

\pf
Given any $b(X)\in J^+_{2m}$. We have an $R_{2m}$-homomorphism
$$\begin{array}{cccc}
\rho_b: & J^+_{2m}\times J_m &\longrightarrow& J^+_{2m}\times J_m,\\[5pt]
 & \big(a(X),a'(X)\big) &\longmapsto& \big(b(X)a(X),b(X)a'(X)\big).
\end{array}$$
Then the image of $\rho_b$ is
$$
 {\rm im}(\rho_b)=I_b\times I'_b = \big\langle b(X)\big\rangle_{R_{2m}}
  \times\big\langle b(X)\big\rangle_{R_{m}}.
$$
Let $(I_b\times I'_b)^{\le\delta}$ denote the set of the words in
$I_b\times I'_b$ whose {\em relative weights}
(the ratio of the Hamming weight to the length) are at most $\delta$, i.e.,
$$\textstyle
(I_b\times I'_b)^{\le\delta}=\big\{(c(X),c'(X))\in I_b\times I'_b
 \,\big|\,\frac{{\rm w}(c(X),c'(X))}{3m}\le \delta\big\}.
$$
Since $X_b$ is a $0$-$1$-variable, the expectation of $X_b$
is just the probability that $X_b=1$, that is,
$$
 {\rm E}(X_b)=\Pr(X_b=1)=\frac{\big|(I_b\times I'_b)^{\le\delta}\big|-1}
 {|I_b\times I'_b|},
$$
where $|S|$ denotes the cardinality of any set $S$. It is clear that
$$
(I_b\times I'_b)^{\le\delta}\subseteq
\bigcup_{w_1,w_2\ge 0,\,w_1+w_2=\lfloor 3m\delta\rfloor}
I_b^{\le\frac{w_1}{2m}}\times I_b'^{\le\frac{w_2}{m}},
$$
where $\lfloor 3m\delta\rfloor$ denotes 
the largest integer which is not larger than $3m\delta$.
Note that $R_{2m}$ ($R_m$ respectively) is a group ring of a cyclic group 
of order $2m$ (order $m$ respectively), and $I_b$ ($I'_b$ respectively) 
is an ideal of $R_{2m}$ ($R_m$ respectively).
By \cite[Theorem 3.3]{FL} and \cite[Remark 3.2]{FL} we obtain that
$$\big|I_b^{\le\frac{w_1}{2m}}\big|\le q^{d_bh_q(\frac{w_1}{2m})},\qquad
 \big|I_b'^{\le\frac{w_2}{m}}\big|\le q^{d_bh_q(\frac{w_2}{m})}.$$
So
\begin{eqnarray*}
\big|(I_b\times\bar I_b)^{\le\delta}\big|
&\le&\sum_{w_1,w_2\ge 0,\,w_1+w_2=\lfloor 3m\delta\rfloor}
q^{d_b h_q(\frac{w_1}{2m})}\cdot q^{d_b h_q(\frac{w_2}{m})}\\
&=&\sum_{w_1,w_2\ge 0,\,w_1+w_2=\lfloor 3m\delta\rfloor}
q^{d_b\big(h_q(\frac{w_1}{2m})+h_q(\frac{w_2}{m})\big)}.
\end{eqnarray*}
Since $h_q(x)$ is a concave function in $[0,1]$
and both $\frac{w_1}{2m}, \frac{w_2}{m}\in [0,1]$
(recall that $\delta<\frac{1}{3}$), we have
$$
h_q\Big(\frac{w_1}{2m}\Big)+h_q\Big(\frac{w_2}{m}\Big)
\le 2 h_q\Big(\frac{\frac{w_1}{2m}+\frac{w_2}{m}}{2}\Big)
=2h_q\Big(\frac{w_1+2w_2}{4m}\Big).
$$
Note that $w_1+w_2=\lfloor 3m\delta\rfloor$ and $\delta<\frac{1}{3}$.
We see that
$$
\frac{w_1+2w_2}{4m}\le\frac{2w_1+2w_2}{4m}
=\frac{3}{2}\delta<\frac{1}{2}\le 1-q^{-1}.
$$
And $h_q(x)$ is increasing in the interval $(0,1-q^{-1})$, so
$$
|(I_b\times I'_b)^{\le\delta}|\le
3m\delta\cdot q^{2d_b h_q(\frac{3\delta}{2})}
\le q^{2d_b h_q(\frac{3\delta}{2})+\log_q m}.
$$
Thus
$$
{\rm E}(X_b)\le \frac{|(I_b\times I'_b)^{\le\delta}|}{|I_b\times I'_b|}
\le q^{-2d_b+2d_b h_q(\frac{3\delta}{2})+\log_q m}.\eqno\qed
$$

From Eqn \eqref{JC_a,a'} of Example \ref{class 11}, 
we have the quasi-cyclic code of index $1\frac{1}{2}$ and co-index $2m$:
\begin{equation}\label{JJC_a,a'}
C_{a,a'}=\big\{(b(X)a(X),b(X)a'(X))\in R_{2m}\times R_m
 \,\big|\,b(X)\in J^+_{2m}\big\},
\end{equation}
where $\big(a(X),a'(X)\big)\in J^+_{2m}\times J_m$.
Since $ J^+_{2m}\times J_m$ is a probability space,
$C_{a,a'}$ is a random code over this probability space,
hence the relative distance $\Delta(C_{a,a'})$ of $C_{a,a'}$ 
is a random variable over the probability space.
We present an estimation of the probability
that $\Delta(C_{a,a'})$ is at most $\delta$.

\begin{lemma}\label{EX}
Let $\delta$ be as in Eqn \eqref{delta} and $C_{a,a'}$ be as in
Eqn \eqref{JJC_a,a'}. Let $\ell_m$ be the minimal degree of
the irreducible divisors of $\frac{X^m-1}{X-1}$ as in Eqn \eqref{l_m}.
Then
$$ \Pr\big(\Delta(C_{a,a'})\le\delta\big)\le
\sum_{j=\ell_m}^{m-1}
 q^{-2j\big(\frac{1}{2}-h_q(\frac{3}{2}\delta)-\frac{\log_q m}{\ell_m}\big)}. 
$$
\end{lemma}

\pf Let $X_b$ for $b(X)\in J^+_{2m}$ be the $0$-$1$-variable in
Eqn \eqref{X_b}. Let
$$X=\sum_{b(X)\in J^+_{2m}} X_b.$$
Then $X$ is a non-negative integer random variable
over the probability space $J^+_{2m}\times J_m$.
By Eqn \eqref{X_b} and Eqn \eqref{JJC_a,a'}, $X$ stands for
the number of $b(X)\in J^+_{2m}$ such that the codeword
$\big(b(X)a(X), b(X)a'(X)\big)$ is non-zero and has Hamming weight
at most $3m\delta$. Thus
$$
 \Pr\big(\Delta(C_{a,a'})\le\delta\big)=\Pr\big(X>0\big).
$$
By a Markov's inequality \cite[Theorem 3.1]{MU},
$\Pr\big(X>0\big)\le {\rm E}(X)$. So we can prove the lemma by
estimating the expectation ${\rm E}(X)$.

By the linearity of the expectation,
${\rm E}(X)=\sum_{b(X)\in J^+_{2m}}{\rm E}(X_b)$.
For any ideal $I$ of $J^+_{2m}$ (we denote it by $I\le J^+_{2m}$), let
$I^*=\big\{b(X)\in I\,\big|\,I_b=I\big\}$, where
$I_b$ is defined in Eqn \eqref{I_b}. That is,
$$
I^*=\big\{b(X)\in I\,\big|\,d_b=\dim I\big\},
$$
where $d_b=\dim I_b$, see Eqn \eqref{I_b}.
It is easy to see that
$ J^+_{2m}=\bigcup_{I\le J^+_{2m}}I^*$, where the subscript
``$I\le J^+_{2m}$'' means that $I$ runs over the ideals contained in $J^+_{2m}$.
By Lemma \ref{number I}, if $0\ne I\le J^+_{2m}$ 
then $\ell_m\le\dim I\le m-1$. So
$$
{\rm E}(X)=\sum_{I\le J^+_{2m}}\sum_{b(X)\in I^*}{\rm E}(X_b)
=\sum_{j=\ell_m}^{m-1}
  \sum_{\mbox{\tiny$\begin{array}{c}I\le J^+_{2m}\\ \dim I=j\end{array}$}}
   \sum_{b(X)\in I^*}{\rm E}(X_b).
$$
For $I\le J^+_{2m}$ with $\dim I=j$, by Lemma \ref{EX_b}
and the fact that $|I^*|\le |I|=q^j$, we get
$$
\sum_{b(X)\in I^*}{\rm E}(X_b)\le \sum_{b(X)\in I^*}
  q^{-2j+2jh_q(\frac{3}{2}\delta)+\log_q m}
  \le q^{-j+2jh_q(\frac{3}{2}\delta)+\log_q m}.
$$
By Lemma \ref{number I} again, the number of
$I\le J^+_{2m}$ with $\dim I=j$ is less that $m^{j/\ell_m}$.
And note that $\log_q m\le\frac{j\log_q m}{\ell_m}$ (as $j\ge\ell_m)$.
So
\begin{eqnarray*}
{\rm E}(X)&\le& \sum_{j=\ell_m}^{m-1} m^{j/\ell_m}
q^{-j+2jh_q(\frac{3}{2}\delta)+\log_q m}\\
&\le& \sum_{j=\ell_m}^{m-1}
 q^{-2j\big(\frac{1}{2}-h_q(\frac{3}{2}\delta)-\frac{\log_q m}{\ell_m}\big)}.
\end{eqnarray*}
The lemma is proved.
\qed

\medskip
By \cite[Lemma 2.6]{BM}, there are positive integers $m_1$, $m_2$, $\cdots$
satisfying that
\begin{equation}\label{m_i}
\gcd(m_i,q)=1,\quad m_i\to\infty,\quad\lim\limits_{i\to\infty}\frac{\log_q m_i}{\ell_{m_i}} =0,
\end{equation}
where $\ell_{m_i}$ is the minimal degree of
the irreducible divisors of $\frac{X^{m_i}-1}{X-1}$ as defined in Eqn \eqref{l_m}.

\begin{theorem}\label{Delta C}
Let $m_1$, $m_2$, $\cdots$ be positive integers
satisfying Eqn \eqref{m_i}.
%such that $\lim\limits_{i\to\infty}\frac{\log_q m_i}{\ell_{m_i}} =0$.
For each $m_i$, let
$C^{(i)}_{a,a'}$ be the random quasi-cyclic code of index $1\frac{1}{2}$
and co-index $2m_i$ as in Eqn~\eqref{JJC_a,a'}.
If $0<\delta<\frac{2}{3}h_q^{-1}(\frac{1}{2})$, then
$$
 \lim\limits_{i\to\infty}\Pr\big(\Delta(C^{(i)}_{a,a'})>\delta\big)=1.
$$
\end{theorem}

\pf
Because of the assumption on $\delta$, we have
$\frac{1}{2}-h_q(\frac{3}{2}\delta)>0$.
By Eqn~\eqref{m_i}, there are a positive real number $\beta$
and an integer $N$ such that
$$\textstyle
 \frac{1}{2}-h_q(\frac{3}{2}\delta)-\frac{\log_q m_i}{\ell_{m_i}} \ge \beta,\qquad \forall~ i>N.
$$
By Lemma \ref{EX},
\begin{eqnarray*}
\lim\limits_{i\to\infty}\Pr\big(\Delta(C^{(i)}_{a,a'})\le\delta\big)
\kern-6pt&\le&\kern-6pt\lim\limits_{i\to\infty}
\sum_{j=\ell_{m_i}}^{m_i-1}q^{-2j\beta}
\le \lim\limits_{i\to\infty}
\sum_{j=\ell_{m_i}}^{m_i-1}q^{-2\ell_{m_i}\beta}\\
&\le&\kern-6pt\lim\limits_{i\to\infty}m_i q^{-2\ell_{m_i}\beta}
=\lim\limits_{i\to\infty}
q^{-2\ell_{m_i}\big(\beta-\frac{\log_q m_i}{2\ell_{m_i}}\big)}.
\end{eqnarray*}
Since $\lim\limits_{i\to\infty}\frac{\log_q m_i}{2\ell_{m_i}}=0$ 
(which implies that $\lim\limits_{i\to\infty}\ell_{m_i}=\infty$),
we obtain that 
$\lim\limits_{i\to\infty}\Pr\big(\Delta(C^{(i)}_{a,a'})\le \delta)=0$.
\qed

Next, we estimate the rate $R(C^{(i)}_{a,a'})$ of the random code $C^{(i)}_{a,a'}$.

\begin{theorem}\label{dim C}
Let $m_1$, $m_2$, $\cdots$ be positive integers
satisfying Eqn \eqref{m_i}.
%such that $\lim\limits_{i\to\infty}\frac{\log_q m_i}{\ell_{m_i}} =0$.
For each $m_i$, let
$C^{(i)}_{a,a'}$ be the random quasi-cyclic code of index $1\frac{1}{2}$
and co-index $2m_i$ as in Eqn~\eqref{JJC_a,a'}. Then
$$
 \lim\limits_{i\to\infty}\Pr\big(\dim C^{(i)}_{a,a'}=m_i-1\big)=1.
$$
\end{theorem}

\pf Let $\frac{X^{m_i}-1}{X-1}=p_1(X)\cdots p_{h_i}(X)$
be the irreducible decomposition in $F[X]$ as in Remark \ref{J_m}.
By Lemma \ref{dC_a,a'} and its notations,
$\dim C^{(i)}_{a,a'}=m_i-1$ if and only if for any $j=1,\cdots,h_i$,
in $F[X]/\langle p_j(X)\rangle\times F[X]/\langle p_j(X)\rangle$ the following
holds:
\begin{equation}\label{mu,mu}
\big(\mu_{2m_i}^{(j)}(a(X)), \;\mu_{m_i}^{(j)}(a'(X)\big)\ne (0,0),
\end{equation}
where $\mu_{2m_i}^{(j)}: J^+_{2m_i}\to F[X]/\langle p_j(X)\rangle$
and $\mu_{m_i}^{(j)}: J_{m_i}\to F[X]/\langle p_j(X)\rangle$
are surjective homomorphisms defined in Remark \ref{J_m}.

Let $d_j=\deg p_j(X)$. Then $F[X]/\langle p_j(X)\rangle$ is a finite field
of cardinality $q^{d_j}$. So
the probability that Eqn~\eqref{mu,mu} holds is equal to
$\frac{q^{2d_j}-1}{q^{2d_j}}=1-q^{-2d_j}$.
Obviously,  the events that Eqn~\eqref{mu,mu}
holds for $j=1,\cdots,h_i$ are randomly independent. Thus
$$
\Pr\big(\dim C^{(i)}_{a,a'}=m_i-1\big)=\prod_{j=1}^{h_i}(1-q^{-2d_j}).
$$
By definition of $\ell_{m_i}$ in Eqn \eqref{l_m},
$\ell_{m_i}\le d_j$ for $j=1,\cdots,h_i$;
hence $h_i\le \frac{m_i-1}{\ell_{m_i}}\le \frac{m_i}{\ell_{m_i}}$. Thus
\begin{eqnarray*}
\Pr\big(\dim C^{(i)}_{a,a'}=m_i-1\big)&\ge&
\big(1-q^{-2\ell_{m_i}}\big)^{\frac{m_i}{\ell_{m_i}}}\\
&=&\big(1-q^{-2\ell_{m_i}}\big)^
  {q^{2\ell_{m_i}}\cdot\frac{m_i}{\ell_{m_i}q^{2\ell_{m_i}}}}.
\end{eqnarray*}
Since $\lim\limits_{i\to\infty}\frac{\log_q m_i}{\ell_{m_i}}=0$
(which implies that $\lim\limits_{i\to\infty}\ell_{m_i}=\infty$),
we see that
$$\lim\limits_{i\to\infty}\frac{m_i}{\ell_{m_i}q^{2\ell_{m_i}}}
 =\lim\limits_{i\to\infty} q^{-\ell_{m_i}\big(2-
  \frac{\log_q m_i}{\ell_{m_i}}+\frac{\log_q\ell_{m_i}}{\ell_{m_i}}\big)}
 =0.
$$
Note that $\big(1-q^{-2\ell_{m_i}}\big)^{q^{2\ell_{m_i}}}>1/4$.
We get that
$$
\lim\limits_{i\to\infty}\Pr\big(\dim C^{(i)}_{a,a'}=m_i-1\big)\ge
\lim\limits_{i\to\infty}(1/4)^{\frac{m_i}{\ell_{m_i}q^{2\ell_{m_i}}}}=1.
\eqno\qed
$$

From Theorem \ref{Delta C} and Theorem \ref{dim C},
we obtain the following at once.

\begin{theorem}\label{a g}
Let $\delta$ be a positive real number
such that $\delta<\frac{2}{3}h_q^{-1}(\frac{1}{2})$.
Then there is a sequence of quasi-cyclic codes $C_i$ of index $1\frac{1}{2}$  over $F$
for $i=1,2,\cdots$ such that the co-index of $C_i$ goes to infinity and
the following hold.

(i)~ $\lim\limits_{i\to\infty}R(C_i)=\frac{1}{3}$;

(ii)~ $\Delta(C_i)>\delta$ for all $i=1,2,\cdots$.
\end{theorem}

For example, if take $q=3$, then
$0.106<\frac{2}{3}h_q^{-1}(\frac{1}{2})<0.107$; so we can take
$\delta=0.106$, and get a sequence $C_1,C_2,\cdots$ of quasi-cyclic ternary codes
of index~$1\frac{1}{2}$ such that the length of $C_i$ goes to infinity,
$R(C_i)\to 1/3$, and $\Delta(C_i)>0.106$ for all $i=1,2,\cdots$.

\section*{Acknowledgements}
The research of the authors is supported by NSFC
with grant numbers 11271005.

\end{document}